\documentclass[12pt]{JHEP3}
\usepackage{amsmath,amssymb}
\newcommand{\be}{\begin{equation}}
\newcommand{\bea}{\begin{eqnarray}}
\newcommand{\eea}{\end{eqnarray}}
\newcommand{\ba}{\begin{array}}
\newcommand{\ea}{\end{array}}
\newcommand{\ee}{\end{equation}}
\def\one{{\hbox{ 1\kern-.8mm l}}}
\newcommand{\Dslash}{\not{\hbox{\kern-4pt $D$}}}
\newcommand{\pdslash}{\not{\hbox{\kern-2pt $\partial$}}}
\newcommand{\cL}{\mathcal{L}} 
\newcommand{\cN}{\mathcal{N}}

\newcommand{\Tr}{\mathrm{Tr}}

\newcommand{\SO}{\mathrm{SO}}

\newcommand{\Comment}[1]{{}}

\def\IZ{{\mathbb Z}}

\def\IR{{\mathbb R}}

\def\caln         {{\cal N}}
\def\calo         {{\cal O}}

\newcommand{\bc}{\begin{center}}
\newcommand{\ec}{\end{center}}
\newcommand{\bmx}{\begin{pmatrix}}
\newcommand{\emx}{\end{pmatrix}}
\newcommand{\nn}{\nonumber}

\newcommand{\del}{\partial}
\newcommand{\half}{\frac{1}{2}}

\newcommand{\tG}{{\tilde G}}

\newcommand{\eref}[1]{Eq.~(\ref{#1})}

\newcommand{\blambda}{{\bar \lambda}}

\newcommand{\hD}{{\hat{D}}}

\newcommand{\tF}{{\tilde F}}
\newcommand{\cG}{{\cal G}}
\newcommand{\tcG}{{\tilde{\cal G}}}

\def\IB{\relax{\rm I\kern-.18em B}}
\def\IC{{\relax\hbox{\kern.3em{\cmss I}$\kern-.4em{\rm C}$}}}
\def\ID{\relax{\rm I\kern-.18em D}}
\def\IE{\relax{\rm I\kern-.18em E}}
\def\IF{\relax{\rm I\kern-.18em F}}
\def\II{\relax{\rm I\kern-.18em I}}
\def\IZ{\relax{\sf Z\kern-.35em Z}}
\def\Id{\relax{1\kern-.32em 1}}
\def\IG{\relax\hbox{$\inbar\kern-.3em{\rm G}$}}
\def\IR{\relax{\rm I\kern-.18em R}}

\normalsize

\title{Higher-derivative 3-algebras}

\author{
Mohsen Alishahiha\,$^a$\,\footnote{email: alishah@ipm.ir}~ 
and Sunil Mukhi\,$^b$\,\footnote{email: mukhi@tifr.res.in}
\\ \it 
${}^a$ School of Physics, Institute for 
Research in Fundamental Sciences (IPM)\\
P.O. Box 19395-5531, Tehran, Iran \\
${}^b$ Tata Institute of Fundamental Research\\
Homi Bhabha Rd, Mumbai 400 005, India}

\abstract{
Starting with the ${\cal N}=8$ supersymmetric Yang-Mills theory on
D2-branes and incorporating higher-derivative corrections to lowest
nontrivial order, we perform a duality to derive the Lorentzian
3-algebra theory along with a set of derivative corrections. We find
that these corrections can be expressed entirely in terms of intrinsic
3-algebra quantities: the 3-bracket and covariant derivatives. Our
analysis is performed for both bosonic and fermionic terms. We
conjecture that the derivative corrections we obtain are relevant
for Euclidean 3-algebra theories as well.
}
\preprint{IPM/P-2008/048, TIFR/TH/08-30}

\begin{document}

\section{Introduction}

It is known that at low energy the world volume modes of $N$ M2-branes
decouple from the eleven-dimensional gravity in the bulk leading to an
${\cal N}=8$ superconformal field theory in three dimensions. This
superconformal theory has an $SO(8)$ R-symmetry which can be
identified with the geometric $SO(8)$ symmetry acting on the eight
transverse directions of the M2-branes. Although we have understood
this theory through its symmetries, it was not clear for over a decade
how to write a model describing three dimensional ${\cal N}=8$
superconformal field theory.

In a series of paper Bagger and Lambert \cite{{Bagger:2006sk},
{Bagger:2007jr},{Bagger:2007vi}} and also Gustavsson
\cite{Gustavsson:2007vu} have constructed an action which is
consistent with all the symmetries of a 3D ${\cal N}=8$ superconformal
field theory; namely it is conformal invariant with 16 supercharges
and has an $SO(8)$ R-symmetry acting on eight scalar fields. Therefore
this model has the potential to describe the world-volume theory of
multiple M2-branes.

This construction relies on the introduction of an algebraic structure
called a ``Lie 3-algebra'' characterized by 4-index structure
constants, ${f^{ABC}}_{D}$ and a bi-linear metric $h^{AB}$. The
structure constants satisfy a fundamental identity which is
essentially a generalization of the Jacobi identity of the Lie
2-algebra. Depending on whether the metric is positive definite or
indefinite we distinguish two cases: Euclidean and Lorentzian
theories\footnote{See
\cite{Gran:2008vi} for an alternative treatment.}. Although the Euclidean
theory, originally proposed by Bagger and Lambert, can only describe a
theory with $SO(4)$ gauge symmetry where $f^{ABCD}=\epsilon^{ABCD}$,
the Lorentzian theory may be written for any classical Lie algebra
\cite{{Gomis:2008uv},{Benvenuti:2008bt},{Ho:2008ei}}.

Even though in the original Lorentzian theories there were potential
ghost-like degrees of freedom, a variant has been proposed that has
been argued to be unitary and describe multiple M2-branes
\cite{{Bandres:2008kj},{Gomis:2008be}}. 
The argument is as follows. One modifies the theory by gauging a shift
symmetry for one of the ``null'' coordinates $X_+^I$ by introducing a
gauge field. The other null coordinate $X_-^I$ is frozen as a result
of the equation of motion of the gauge field.  Therefore the resultant
theory is manifestly ghost-free. Indeed, using the Higgs mechanism of
Ref.\cite{Mukhi:2008ux} it was shown \cite{Ho:2008ei,Honma:2008un}
that the theory reduces to maximally supersymmetric Yang-Mills in
three dimensions whose gauge coupling is the vev of the scalar
field. This result indicates that the ghost free Lorentzian theory is
closely related to SYM theory. However in \cite{Ezhuthachan:2008ch} it
was shown that starting from maximally supersymmetric 3D Yang-Mills
theory and using a duality transformation due to de Wit, Nicolai and
Samtleben
\cite{{Nicolai:2003bp},{deWit:2003ja},{deWit:2004yr}}, one can
directly obtain the ghost-free Lorentzian 3-algebra
theory\footnote{The same mechanism was subsequently used to derive
globally $N=8$ supersymmetric actions from
supergravity\cite{Bergshoeff:2008ix}.}. Since it can be derived from
SYM, the final theory is manifestly equivalent to it on-shell. Though
it does have enhanced R-symmetry as well as superconformal symmetry
off-shell, it is the D2-brane theory on-shell for any {\it finite} vev
of the gauge-singlet scalar field.

On the other hand at higher orders in $\alpha'$ the world-volume
theory of multiple D2-branes is believed to be described by some
non-Abelian generalization of the DBI action. Therefore, one would
expect that the 3-algebra theories just represent the lowest order of
the full effective action describing the world-volume of multiple
M2-branes.  Therefore it should be interesting to study non-linear
corrections to 3-algebra theories. One straightforward approach is to
consider these corrections in the context of Lorentzian 3-algebras,
where as indicated above they should be {\it derivable} from the SYM
theory. 

Accordingly, in this article we extend the considerations of
\cite{Ezhuthachan:2008ch} when higher-derivative corrections are
taken into account. More precisely starting with the ${\cal N}=8$
supersymmetric Yang-Mills theory on D2-branes and incorporating
higher-derivative corrections to lowest nontrivial order, we perform a
duality to derive the Lorentzian 3-algebra theory along with a set of
derivative corrections given by non-Abelian $F^4$ terms
\cite{Tseytlin:1997csa}. We will show that these corrections assemble
themselves neatly into the basic objects of a 3-algebra, namely the
3-bracket and covariant derivatives. This holds for both bosonic and
fermionic terms and we provide explicit forms for the leading
correction in both cases. 

Finally we conjecture that the derivative corrections we have obtained
here, being independent of the details of the 3-algebra, should be
relevant for Euclidean 3-algebra theories as well. This conjecture in
principle enlarges the potential applicability of the results in this
paper to a wider class of 3-algebras beyond the Lorentzian-signature
ones. However, because the 3-bracket for us is totally antisymmetric,
our results can be immediately generalized at this stage only to
maximally supersymmetric (${\cal N}=8$) Euclidean 3-algebras, of which
the sole example is the Bagger-Lambert $A_4$
theory\cite{Bagger:2007vi}. It may be possible in the future to extend
these considerations to 3-algebra theories with lower supersymmetry
such as those discussed in Refs.\cite{Schnabl:2008wj, Bagger:2008se}
(see also \cite{Bergshoeff:2008bh}).

The rest of the paper is organized as follows. In section two we will
set our notation by reviewing the construction of
Ref.\cite{Ezhuthachan:2008ch}. In section three we will extend the
results to incorporate bosonic non-Abelian $F^4$ terms and the
corresponding scalar terms. In
section four we discuss some general features of these higher order
corrections. In section five we obtain the $SO(8)$ covariant fermionic
terms to the same order in $\alpha'$. Finally we present a conjecture
and our conclusions.

\section{Review}
  
We would like to consider the maximally supersymmetric interacting
super Yang-Mills Lagrangian in 2+1 dimensions based on an arbitrary
Lie algebra ${\cal G}$ whose bosonic action in leading order is given
by:
\begin{equation}\label{sym}
\begin{split}
  \cL = \Tr\left( -\frac{1}{4g^{2}_{YM}}
F_{\mu\nu}F^{\mu\nu} - \frac{1}{2}D_\mu
    X^iD^\mu X^i - \frac{g_{YM}^2}{4}[X^i,X^j][X^j,X^i]\right)\;,
\end{split}
\end{equation}
Here $A_\mu$ is a gauge connection on ${\cal G}$. The field
strength and the covariant derivatives are defined as:
\begin{equation}
\label{fscd}
  F_{\mu\nu} = \partial_\mu A_\nu - \partial_\nu A_\mu - 
  [A_\mu,A_\nu]\qquad \textrm{and}\qquad D_\mu = \partial_\mu -
  [A_\mu,\cdot\;]\;.
\end{equation} 
The $X^i$s are seven matrix valued scalar fields transforming as
vectors under the $\SO(7)$ R-symmetry group. 

In \cite{Ezhuthachan:2008ch} it was shown that this Lagrangian
can be brought to the form of the Lorentzian Bagger-Lambert or
3-algebra field theory proposed in
\cite{Gomis:2008uv,Benvenuti:2008bt,Ho:2008ei}, or more precisely to
the ``gauged'' version of the above theory described in
\cite{Bandres:2008kj,Gomis:2008be}. 
Here we first review the results of \cite{Ezhuthachan:2008ch}.

We proceed by introducing two new fields $B_{\mu}$ and $\phi$ that are
adjoints of ${\cal G}$. In terms of these new fields the
dNS duality transformation
\cite{Nicolai:2003bp,deWit:2003ja,deWit:2004yr} is the replacement:
\begin{equation}\label{dual}
\Tr\left( -\frac{1}{4 g_{YM}^2}
  F^{\mu\nu}F_{\mu\nu}\right) \rightarrow \Tr \left(
  \frac{1}{2}\epsilon^{\mu\nu\lambda}B_\mu F_{\nu\lambda}
  -\frac{1}{2}\left(D_{\mu}\phi- g_{YM} B_{\mu} \right)^2\right)\;.
\end{equation}
We see that in addition to the gauge symmetry $\cG$, the new action
has a noncompact Abelian gauge symmetry that we can call $\tcG$,
which has the same dimension as the original gauge group
$\cG$. This symmetry consists of the transformations:
\be
\label{ncgt}
 \delta \phi = g_{YM}M \;,\qquad \delta B_\mu = D_\mu M\;,
\ee
where $M(x)$ is an arbitrary matrix, valued in the adjoint of ${\cal
G}$. Clearly $B_{\mu}$ is the gauge field for the shift symmetries
$\tilde{\cal G}$. Note that both in \eref{dual} and \eref{ncgt}, the
covariant derivative $D_\mu$ is the one defined in \eref{fscd}.

If one chooses the gauge $D^\mu B_\mu=0$ to fix the shift symmetry,
the degree of freedom of the original Yang-Mills gauge field $A_\mu$
can be considered to reside in the scalar $\phi$. In this sense one
can think of $\phi$ as morally the dual of the original
$A_\mu$~\cite{Nicolai:2003bp,deWit:2003ja,deWit:2004yr}.
Alternatively we can choose the gauge $\phi=0$, in which case the same
degree of freedom resides in $B_\mu$.  The equivalence of the RHS to
the LHS of \eref{dual} can be conveniently seen by going to the latter
gauge. Once $\phi=0$ then $B_\mu$ is just an auxiliary field and one
can integrate it out to find the usual YM kinetic term for
$F_{\mu\nu}$.

We can now proceed to study the dNS-duality transformed of the bosonic
sector of $\mathcal N =8$ Yang-Mills theory. Its Lagrangian is:
\begin{equation}\label{D2}
\begin{split}
  \cL = &\Tr\left(\frac{1}{2}
      \epsilon^{\mu\nu\lambda}B_{\mu}F_{\nu\lambda} -
    \frac{1}{2} \left(D_{\mu}\phi -g_{YM}B_{\mu}\right)^{2}
    -\frac{1}{2}D_\mu X^i D^\mu X^i \right.  \\ 
\qquad &\left.\;\;\;\;\;\;\;-
    \frac{g_{YM}^2}{4}[X^i,X^j][X^j,X^i] \right)\;.
\end{split}
\end{equation}

The gauge-invariant kinetic terms for the eight scalar fields have a
potential $\SO(8)$ invariance, which can be exhibited as follows. First
rename $\phi(x) \to X^8(x)$. Then the scalar kinetic terms become
$-\half \hD_\mu X^I \hD^\mu X^I$, where:
\bea
\label{covder}
\hD_\mu X^i &=& D_\mu X^i = \del_\mu X^i -  [A_\mu,X^i],
\quad i=1,2,\ldots,7\nn\\
\hD_\mu X^8 &=& D_\mu X^8 - g_{YM}B_\mu = \del_\mu X^8 -  [A_\mu,X^8]
- g_{YM} B_\mu\;.
\eea
Defining the constant 8-vector:
\be
\label{gvec}
g_{YM}^I = (0,\ldots,0,g_{YM})\;,\quad I=1,2,\ldots,8\;,
\ee 
the covariant derivatives can together be written:
\be
\label{newcovder}
\hD_\mu X^I = D_\mu X^I - g_{YM}^I B_\mu\;.
\ee

One can now uniquely write the SYM action in a form that is
$\SO(8)$-invariant under transformations that rotate both the fields
$X^I$ and the coupling-constant vector $g_{YM}^I$:
\begin{equation}\label{so8} 
\begin{split}
  \cL =&  \Tr\Big( \frac{1}{2}\epsilon^{\mu\nu\lambda}B_\mu F_{\nu\lambda}
   - \frac{1}{2}\hat{D}_\mu X^I \hat{D}^\mu X^I  \\
&- \frac{1}{12} \left( g_{YM}^I [X^J,X^K] +
    g_{YM}^J[X^K,X^I]  + g_{YM}^K[X^I,X^J]\right)^2 \Big)\;.
  \end{split}
\end{equation}

The final step is to replace $g_{YM}^I$ by a scalar field $X_+^I$ that
is constrained to be a constant\footnote{Flux quantization in the
original theory implies the matrix-valued scalars have a periodicity
$X^I\sim X^I + X_+^I \II$. We thank Juan Maldacena for emphasizing
this to us.}. This proceeds as described in
\cite{Ezhuthachan:2008ch} and we will describe it again in the
following section where we address higher-derivative terms. The
fermionic contributions also must be added, and these too will be
described in what follows.

\section{$F^4$ terms}

The aim of this section is to redo the procedure of the previous
section for subleading terms of the three dimensional theory. The
subleading terms consist of $F^4$ with four derivative interactions of
the scalar fields. To find the explicit terms we note that the leading
order terms in the action can be found from reduction of the ten
dimensional pure gauge Yang-Mills theory.  Therefore to get the higher
derivative terms for the three dimensional theory we will start from
ten dimensional $F^4$ terms given by
\cite{Tseytlin:1997csa}\footnote{We are using units in which
$\alpha'=\frac{1}{2\pi}$.}
\be
\label{ffourterms}
\begin{split}
L^{(10)}=\frac{1}{12}{\rm Tr} \bigg{[}&F_{MN}F_{RS}F^{MR}F^{NS}
+\frac{1}{2}F_{MN}F^{NR}F_{RS}F^{SM}
-\frac{1}{4}F_{MN}F^{MN}F_{RS}F^{RS}\cr &
-\frac{1}{8}F_{MN}F_{RS}F^{MN}F^{RS}
\bigg{]},
\end{split}
\ee
where $M,N,R,S=0,\cdots,9$. The aim is now to reduce this action to
three dimensions. To do that we decompose the indices to
$\mu,\nu,\rho,\sigma=0,1,2$ and $i,j,k,l=1,\cdots, 7$. Then the
Yang-Mills plus $F^4$ terms lead to the following Lagrangian:
\be
L^{(4)}=L^{(2)}+\sum_{i=1}^6{\rm Tr}\; L^{(4)}_i,
\ee
where
\be
\begin{split}
L^{(2)} &= -\frac{1}{4g_{YM}^2}F_{\mu\nu}F^{\mu\nu}\cr
L^{(4)}_1&=\frac{1}{12g_{YM}^4}\bigg{[}F_{\mu\nu}F_{\rho\sigma}
F^{\mu\rho}F^{\nu\sigma}+\frac{1}{2}F_{\mu\nu}F^{\nu\rho}F_{\rho\sigma}
F^{\sigma\mu}
-\frac{1}{4}F_{\mu\nu}F^{\mu\nu}F_{\rho\sigma}F^{\rho\sigma}\cr
&-\frac{1}{8}F_{\mu\nu}F_{\rho\sigma}F^{\mu\nu}F^{\rho\sigma}\bigg{]}
\end{split}
\ee
\be
\begin{split}
L^{(4)}_2=\frac{1}{12g_{YM}^2}&\bigg{[}F_{\mu\nu}\;D^\mu
X^i\;F^{\rho\nu}\;D_\rho X^i +F_{\mu\nu}\;D_\rho
X^i\;F^{\mu\rho}\;D^\nu X^i-2 F_{\mu\rho}\;F^{\rho\nu}\;D^\mu
X^i\;D_\nu X^i\cr -&2F_{\mu\rho}\;F^{\rho\nu}\; D_\nu X^i\; D^\mu
X^i-F_{\mu\nu}\;F^{\mu\nu}\;D^\rho X^i\;D_\rho X^i -\frac{1}{2}
F_{\mu\nu}\;D_\rho X_i\;F_{\mu\nu}\;D_\rho X_i \bigg{]}\cr
-&\frac{1}{12}\left( \frac{1}{2}F_{\mu\nu}\;F^{\mu\nu}\;X^{ij}\;
X^{ij} +\frac{1}{4}F_{\mu\nu}\;X^{ij}\;F^{\mu\nu}\;X^{ij}\right)
\end{split}
\ee
\be
\begin{split}
L^{(4)}_3=-\frac{1}{6}\bigg{(}&D^\mu X^i\;D^\nu X^jF_{\mu\nu}+D^\nu X^j\;
F_{\mu\nu}\;D^\mu X^i+F_{\mu\nu}\;D^\mu X^i\;D^\nu X^j\bigg{)}\;X^{ij}
\end{split}
\ee
\be
\begin{split}
L^{(4)}_4=\frac{1}{12}&\bigg{[}D_{\mu}X^i\;D_{\nu}X^j\;D^\nu X^i\;D^\mu X^j
+D_{\mu}X^i\;D_{\nu}X^j\;D^\mu X^j\;D^\nu X^i\cr
&+D_{\mu}X^i\;D_{\nu}X^i\;D^\nu X^j\;D^\mu X^j 
-D_{\mu}X^i\;D^{\mu}X^i\;D_\nu X^j\;D^\nu X^j\cr
&-\frac{1}{2}D_{\mu}X^i\;D_{\nu}X^j\;D^\mu X^i\;D^\nu X^j\bigg{]}
\end{split}
\ee
\be
\begin{split}
L^{(4)}_5=\frac{g_{YM}^2}{12}&\bigg{[}X^{kj}\; D_\mu
X^k\;X^{ij}\;D^\mu X^i+X^{ij}\;D_\mu X^{k}\;X^{ik}
\;D^\mu X^j\cr 
&-2X^{kj}\;X^{ik}\;D_\mu X^j\;D^\mu X^i-2X^{ki}\;X^{jk}\;D_\mu
X^j\;D^\mu X^i\cr &-X^{ij}\;X^{ij}\;D_\mu X^k\;D^\mu
X^k-\frac{1}{2}X^{ij}\;D_\mu X^k\;X^{ij}\;D^\mu X^k\bigg{]}
\end{split}
\ee
\be
\begin{split}
L^{(4)}_6=\frac{g_{YM}^4}{12}\bigg{[}&X^{ij}X^{kl}X^{ik}X^{jl}
+\frac{1}{2}X^{ij}X^{jk}X^{kl}
X^{li} -\frac{1}{4}X^{ij}X^{ij}X^{kl}X^{kl}\cr
&-\frac{1}{8}X^{ij}X^{kl}X^{ij}X^{kl}\bigg{]}
\end{split}
\ee

Following the previous section the aim is to rewrite the above
Lagrangian in terms of the new fields, $B_\mu,X^8$ such that the
obtained Lagrangian will be manifestly $SO(8)$ invariant.  It is
useful to proceed in two steps. First we simply rewrite the Lagrangian
in terms of the Poincare dual field strength defined by:
\be
\tF_\mu \equiv \half\epsilon_{\mu\nu\lambda}F^{\nu\lambda}
\ee
Note that in our conventions (with a $(-++)$ metric), the inverse
transformation is $F_{\mu\nu}= -\epsilon_{\mu\nu\lambda}\tF^\lambda$.
Later we will replace $\tF$ by an independent field $B_\mu$ that will
be subjected to constraints via the equations of motion, leading back
to the original action.

Replacing $F_{\mu\nu}$ in terms of $\tF_\mu$ everywhere in the
preceding Lagrangian, we end up with:
\bea
&&L^{(2)} + L_1^{(4)} + L_2^{(4)} + L_3^{(4)}=
{\rm Tr}\bigg{[}\frac{1}{2g_{YM}^2} \tF_\mu \tF^\mu
+\frac{1}{12 g_{YM}^4}\bigg{(}\tF_\mu \tF^\mu \tF_\nu
\tF^\nu+\frac{1}{2}\tF_\mu \tF_\nu \tF^\mu \tF^\nu\bigg{)}
\nn\\
&&+\frac{1}{12 g_{YM}^2}\bigg{(}2 \tF^\mu \tF_\nu\; D^\nu X^i\; D_\mu
X^i -2\tF^\mu \tF_\mu\; D_\nu X^i\; D^\nu X^i +2 \tF^\mu \tF^\nu\;
D_\mu X^i\; D_\nu X^i\cr 
&&\;\;\;\;\;\;\;\;\;\;\;\;+\tF^\mu\; D^\nu
X^i\;\tF_\nu\; D_\mu X^i- \tF^\mu\; D^\nu X^i\; \tF_\mu\; D_\nu X^i
+\tF^\mu\; D_\mu X^i\; \tF^\nu\; D_\nu X^i\bigg) \nn\\
&&\;\;\;\;\;\;\;\;\;\;\;\;+\frac{1}{12}\bigg(\tF^\mu\;\tF_\mu\; X^{ij}\; X^{ij}
+\frac{1}{2}\tF^\mu\; X^{ij}\;\tF_\mu\; X^{ij}\bigg{)}\cr
&&+\frac{1}{6}\epsilon_{\rho\mu\nu}\bigg{(}\tF^\rho\;D^\mu
X^i\;{D}^\nu X^j +{D}^\nu X^j\;{\tF}^\rho \;{D}^\mu X^i+{D}^\mu
X^i\;{D}^\nu X^j\;{\tF}^\rho
\bigg{)}X^{ij}\bigg{]}\nn\\
\label{BBBF}
\eea
Here we have written only the terms involving $\tF$, as the remaining
ones $L_4^{(4)},L_5^{(4)},L_6^{(4)}$ are obviously unaffected by our
substitution.

Let us now perform a dNS duality, as in the previous section, but in
the presence of the above higher-derivative corrections.  Introducing
again an independent 1-form (matrix-valued) field $B_\mu$, it is easy
to see that the above action can be replaced with one where $\tF$
appears only in the Chern-Simons interaction $\tF_\mu B^\mu$:
\bea
&&L^{(2)} + L_1^{(4)} + L_2^{(4)} + L_3^{(4)}=
{\rm Tr}\bigg{[}\tF_\mu B^\mu - \frac{g_{YM}^2}{2} B_\mu B^\mu
\nn\\
&& +\frac{g_{YM}^4}{12}\bigg{(}B_\mu B^\mu B_\nu
B^\nu+\frac{1}{2}B_\mu B_\nu B^\mu B^\nu\bigg{)}\\
&&+\frac{g_{YM}^2}{12}\bigg{(}2 B^\mu B_\nu\; D^\nu X^i\; D_\mu
X^i -2B^\mu B_\mu\; D_\nu X^i\; D^\nu X^i +2 B^\mu B^\nu\;
D_\mu X^i\; D_\nu X^i\nn\\ 
&&\;\;\;\;\;\;\;\;\;\;\;\;+B^\mu\; D^\nu
X^i\;B_\nu\; D_\mu X^i- B^\mu\; D^\nu X^i\; B_\mu\; D_\nu X^i
+B^\mu\; D_\mu X^i\; B^\nu\; D_\nu X^i\bigg) \nn\\
&&\;\;\;\;\;\;\;\;\;\;\;\;+\frac{g_{YM}^4}{12}
\bigg(B^\mu\;B_\mu\; X^{ij}\; X^{ij}
+\frac{1}{2}B^\mu\; X^{ij}\;B_\mu\; X^{ij}\bigg{)}\nn\\
&&+\frac{g^2_{YM}}{6}\epsilon_{\rho\mu\nu}\bigg{(}B^\rho\;D^\mu
X^i\;{D}^\nu X^j +{D}^\nu X^j\;{B}^\rho \;{D}^\mu X^i+{D}^\mu
X^i\;{D}^\nu X^j\;{B}^\rho
\bigg{)}X^{ij}\bigg{]}\nn
\label{BBBFtwo}
\eea
To show that this substitution is correct, simply integrate out the
field $B$ order by order (truncating at quartic order, since that is
all the input we had to start with) using its equation of motion.  It
is easy to check that this brings the above Lagrangian to the form:
\be 
L^{(2)}+{\rm Tr}(L^{(4)}_1+L^{(4)}_2+L^{(4)}_3)+\calo(F^6).
\ee
We now use this form, depending on the new field $B_\mu$, to rewrite
the Lagrangian in an $SO(8)$ invariant way. For this, introduce the
field $X^8$ and replace $B_\mu$, everywhere it occurs, by
$-\frac{1}{g_{YM}}(D_\mu X^8-g_{YM} B_\mu)$. There is now a shift
symmetry as in \eref{ncgt} using which one can set $X^8=0$ and we get
back to the above action. The utility of this transformation will be
that in more general gauges, $X^8$ can carry the dynamical degree of
freedom. 

As explained in Eqs.(\ref{covder}),(\ref{gvec}),(\ref{newcovder}), it
is useful to write the coupling constant formally as an 8-vector,
since this allows us to express all the covariant derivatives in a
unified manner as $\hat{D}_\mu X^I, I=1,2,\cdots,8$. Then under the
above replacement,
\eref{BBBF} becomes\footnote{Using integration by parts 
and cyclicity of the trace one can show that the
$\tilde{F}^\mu D_\mu X^8$ term vanishes.}:
\bea
&&{\rm Tr}\bigg{[}\frac{1}{2}\epsilon^{\mu\nu\rho}B_\mu
F_{\nu\rho}-\frac{1}{2}\hat{D}_\mu X^8\hat{D}^\mu X^8
\nn\\[1mm]
&&+\frac{1}{12}\bigg{(}\hat{D}_\mu X^8\hat{D}^\mu X^8\hat{D}_\nu
X^8\hat{D}^\nu X^8 +\frac{1}{2}\hat{D}_\mu X^8\hat{D}_\nu
X^8\hat{D}^\mu X^8\hat{D}^\nu X^8\bigg{)}\nn\\[1mm]
&&+\frac{1}{12}\bigg{(}2\hat{D}^\mu X^8\;\hat{D}_\nu X^8\;\hat{D}^\nu
X^i \;\hat{D}_\mu X^i -2\hat{D}^\mu X^8\;\hat{D}_\mu X^8\;\hat{D}_\nu
X^i\; \hat{D}^\nu X^i\nn\\[1mm]
&&\;\;\;\;\;\;\;\;\;+2\hat{D}^\mu
X^8\;\hat{D}^\nu X^8\;\hat{D}_\mu X^i\; \hat{D}_\nu X^i
+\hat{D}^\mu X^8\;\hat{D}^\nu X^i\;\hat{D}_\nu
X^8\;\hat{D}_\mu X^i\nn\\[1mm] 
&&\;\;\;\;\;\;\;\;\;
-\hat{D}^\mu X^8\;\hat{D}^\nu X^i\; \hat{D}_\mu
X^8\;\hat{D}_\nu X^i 
+\hat{D}^\mu
X^8\;\hat{D}_\mu X^i\;\hat{D}^\nu X^8\;\hat{D}_\nu X^i \bigg)\\[1mm]
&&\;\;\;\;\;\;\;\;\;+\frac{g_{YM}^2}{12}\bigg(
\hat{D}^\mu X^8\;\hat{D}_\mu X^8\;
X^{ij}\; X^{ij} +\frac{1}{2}\hat{D}^\mu X^8\;
X^{ij}\;\hat{D}_\mu X^8\; X^{ij}\bigg{)}\nn\\[1mm]
&&+\frac{g_{YM}}{6}\epsilon_{\rho\mu\nu}\bigg{(}\hat{D}^\rho
X^8\;\hat{D}^\mu X^i\;\hat{D}^\nu X^j +\hat{D}^\nu X^j\;\hat{D}^\rho
X^8\;\hat{D}^\mu X^i+\hat{D}^\mu X^i\;\hat{D}^\nu X^j\;\hat{D}^\rho
X^8
\bigg{)}X^{ij}\bigg{]}\nn
\eea

It is now straightforward, though a little messy, to
see that the leading order terms given in equation (\ref{sym}) plus
$\sum_{i=1}^6{\rm Tr}\; L^{(4)}_i$ can be written in the $SO(8)$
invariant terms as follows:
\bea
\label{bosonfinal}
&&{\rm Tr}\bigg{[}\frac{1}{2}\epsilon^{\mu\nu\rho}B_\mu
F_{\nu\rho}-\frac{1}{2}\hat{D}_\mu X^I\hat{D}^\mu X^I 
\nn\\[1mm] 
&&+\frac{1}{12}\bigg{(}\hat{D}_{\mu}X^I\;\hat{D}_{\nu}X^J\;\hat{D}^\nu
X^I\;\hat{D}^\mu X^J+\hat{D}_{\mu}X^I
\;\hat{D}_{\nu}X^J\;\hat{D}^\mu X^J\;\hat{D}^\nu X^I\nn\\[1mm]
&&\;\;\;\;\;\;\;\;\;+\hat{D}_{\mu}X^I\;\hat{D}_{\nu}X^I\;\hat{D}^\nu
X^J\;\hat{D}^\mu X^J-\hat{D}_{\mu}X^I
\;\hat{D}^{\mu}X^I\;\hat{D}_\nu X^J\;\hat{D}^\nu X^J\nn\\[1mm]
&&\;\;\;\;\;\;\;\;\;-\frac{1}{2}
\hat{D}_{\mu}X^I\;\hat{D}_{\nu}X^J\;\hat{D}^\mu
X^I\;\hat{D}^\nu X^J\bigg{)}\nn\\[1mm]
&&+\frac{1}{12}\bigg{(}\frac{1}{2}X^{LKJ}\; 
\hD_\mu X^K\;X^{LIJ}\;\hD^\mu
X^I+\frac{1}{2}X^{LIJ}\;\hD_\mu X^{K}\;X^{LIK}
\;\hD^\mu X^J\nn\\[1mm] 
&&\;\;\;\;\;\;\;\;\;\;-X^{LKJ}\;X^{LIK}\;\hD_\mu X^J\;\hD^\mu
X^I-X^{LKI}\;X^{LJK}\;\hD_\mu X^J\;\hD^\mu X^I\nn\\[1mm]
&&\;\;\;\;\;\;\;\;\;\;
-\frac{1}{3}X^{LIJ}\;X^{LIJ}\;\hD_\mu X^K\;\hD^\mu
X^K-\frac{1}{6}X^{LIJ}\;\hD_\mu X^K\;X^{LIJ}\;\hD^\mu X^K\bigg{)}\nn\\[1mm]
&&\;\;\;\;\;\;\;\;\;\;
-\frac{1}{6}\epsilon_{\rho\mu\nu}\hat{D}^\rho X^I\;\hat{D}^\mu
X^J\;\hat{D}^\nu X^K X^{IJK} -V(X)\bigg]
\label{act1}
\eea
where 
\be
\label{threeprod}
X^{IJK}= g_{YM}^I [X^J,X^K] + g_{YM}^J[X^K,X^I]  + g_{YM}^K[X^I,X^J]
\ee
Here $V(X)$ is the potential:
\bea
\label{act2}
V(X)&=& \frac{1}{12}
{X^{IJK}} {X^{IJK}}
+\frac{1}{9\times
12}\bigg{[}X^{NIJ}X^{NKL}X^{MIK}X^{MJL}\\[1mm]
&+&\frac{1}{2}X^{NIJ}X^{MJK}X^{NKL}
X^{MLI} -\frac{1}{4}X^{NIJ}X^{NIJ}X^{MKL}X^{MKL}\cr
&-&\frac{1}{8}X^{NIJ}X^{MKL}X^{NIJ}X^{MKL}\bigg{]}\nn
\eea

Once we have $SO(8)$ covariance, we are free to replace the fixed
vector of coupling constants $g_{YM}^I$ by any arbitrary vector with
the same modulus. The last step is to replace these constants by a
set of scalar fields $X_+^I$ and introduce another scalar $X_-^I$ as
well as a gauge field $C^{\mu,I}$ with the kinetic term:
\be
(C^{\mu\,I}-\del^\mu X_-^I)\del^\mu X_+^I
\ee
As explained in Refs.\cite{Gomis:2008be,Ezhuthachan:2008ch}, this
has the effect of constraining the vector $X_+^I$ to be an arbitrary
constant which we can then identify with $g_{YM}^I$.

Thus the final form of our derivative-corrected action is as in
Eqs.(\ref{act1}) and (\ref{act2}), with the covariant derivatives
replaced by:
\be
{\hat D}_\mu X^I = \del_\mu - [A_\mu,X^I] - B_\mu X_+^I
\ee
and the commutator terms \eref{threeprod} replaced by the Lorentzian
3-algebra triple product:
\be
\label{newthreeprod}
X^{IJK}= X_+^I [X^J,X^K] + X_+^J[X^K,X^I]  + X_+^K[X^I,X^J]
\ee
This must be supplemented, of course, with fermionic terms as well as
gauge-fixing terms for the various local symmetries. We will discuss
the fermions in detail in a subsequent  section.

To summarize, in this section we have used dNS duality to re-write the
three dimensional $\caln=8$ supersymmetric Yang-Mills theory,
including the first nontrivial derivative corrections, in a form which
is manifestly $SO(8)$ invariant. We now turn to a discussion of the
generality of this result.

\section{Generality of the result and higher order terms}

Encouraged by what we have found, we would like in this section to ask
how general the result is. Is it true that to any order, the
derivative correction computed for $\cN=8$ SYM in 3d can be
re-expressed in $SO(8)$ invariant form? Specifically we wish to
understand whether achieving $SO(8)$ invariance depends on the
specific combination of $F^4$ terms appearing in
\eref{ffourterms}. If this is not the case, in other words 
if enhanced $SO(8)$ is generically present for any higher order $F^n$
terms that one can think of writing down in 10d, then it would not be
such a miracle. But in fact, as we will see below, $SO(8)$ enhancement
does {\it not} hold for generic higher-order corrections. The specific
combination occurring in \eref{ffourterms}, which arises from string
theory, is essential for the result that we found in order $F^4$, and
a similar situation is expected to hold in higher orders.

Instead of considering the most general case, we will find it
illuminating to start with a simplified approach. Consider an Abelian
SYM theory in 10d. Let us now postulate a generic quartic correction
to the 10d Lagrangian, namely:
\be
L^{(4)}_{10d} = 
\lambda_1\, F_{AB}F^{AB}F_{CD}F^{CD}
+ \lambda_2\, F^A_{~B} F^B_{~C} F^C_{~D} F^D_{~A}
\ee
where we have put arbitrary coefficients in front of the two possible
quartic terms. (In this section we set $g_{YM}=1$ for notational
simplicity.) After reducing to 3d, the field strength terms can be
dualized to 1-forms as before (using $\tF_\mu =
\half\epsilon_{\mu\nu\lambda}F^{\nu\lambda}$) and we find:
\be
\label{lfourgauge}
L^{(4)}_{\rm gauge} = 
\left(4\lambda_1 + 2\lambda_2\right) 
\tF_{\mu}\tF^{\mu}\tF_{\nu}\tF^{\nu}
\ee
Note that two different tensor structures in 10d have reduced to the
same one in 3d. This is because of the duality between 1-forms and
2-forms in 3d. On the other hand, the terms involving $\del X$ are
found to be:
\bea
L^{(4)}_{\del X}
\label{lfourdelx}
&=&-\left(8\lambda_1 + 4\lambda_2\right)\del_\mu X^i \del^\mu X^i
\,\tF_\nu \tF^\nu + 4\lambda_2\, \del_\mu X^i \del_\nu X^i \tF^\mu
\tF^\nu\nn\\
&&+~4\lambda_1\, \del_\mu X^i \del^\mu X^i \del_\nu X^j \del^\nu X^j
+ 2\lambda_2\, \del_\mu X^i \del_\nu X^i \del^\mu X^j \del^\nu X^j
\eea
where as usual the indices $i,j=1,2,\cdots,7$. For the Abelian case
Eqns.(\ref{lfourgauge}),(\ref{lfourdelx}) make up the whole reduced
action to this order, since commutator terms are absent.

Now let us ask if the above expression has $SO(8)$ invariance after
performing dNS duality. To quartic order this duality merely replaces
$\tF_\mu$ everywhere in the quartic terms by $B_\mu$ (as we will see,
this is not the the case from order 6 onwards). After that, we
replace $B_\mu$ by $-\del_\mu X^8$. The result for the quartic action
$L^{(4)}_{3d} = L^{(4)}_{\rm gauge} + L^{(4)}_{\del X}$ is:
\bea
L^{(4)}_{3d} &=&
\left(4\lambda_1 + 2\lambda_2\right) 
\del_\mu X^8 \del^\mu X^8 \del_\nu X^8 \del^\nu X^8
-\left(8\lambda_1 + 4\lambda_2\right)\del_\mu X^i \del^\mu X^i
\del_\nu X^8\del^\nu X^8\nn\\ 
&&+ 4\lambda_2\, \del_\mu X^i \del_\nu X^i \del^\mu X^8 \del^\nu X^8
+~4\lambda_1\, \del_\mu X^i \del^\mu X^i \del_\nu X^j \del^\nu
X^j\nn\\ 
&&+ 2\lambda_2\, \del_\mu X^i \del_\nu X^i \del^\mu X^j \del^\nu X^j
\eea
Now it is easy to see that the above action is equal to the $SO(8)$
invariant combination:
\be
4\lambda_1\, \del_\mu X^I \del^\mu X^I \del_\nu X^J \del^\nu
X^J + 2\lambda_2\, \del_\mu X^I \del_\nu X^I \del^\mu X^J \del^\nu
X^J
\ee
where $I,J=1,2,\cdots,8$, but only if the following constraint is
satisfied:
\be
\lambda_2 = -4\lambda_1
\ee
Without this constraint, $L_{3d}^{(4)}$ cannot be recast in 
$SO(8)$ invariant form.

In light of this simple computation, we may go back to the previous
section and see if that computation, specialized to the Abelian case,
satisfies our constraint above. Once we treat all $F$'s as commuting,
we find that the four coefficients in \eref{ffourterms} collapse to
two independent coefficients corresponding to $\lambda_1 =
-\frac{1}{32}$ and $\lambda_2 =
\frac18$. Therefore the above constraint is satisfied. This explains
why we found $SO(8)$ invariance in the previous section and makes it
clear that this was crucially dependent on using the corrections that
arise in string theory (which evidently ``knows'' about this
constraint) and would not have worked for generic correction terms.

In fact, for the Abelian case it is an old result 
\cite{Townsend:1995af,Schmidhuber:1996fy} that $SO(8)$ invariance can be
obtained for the full DBI action by performing a duality. We summarize
that argument here after translating it into our conventions for ease
of comparison, and presenting in the more ``modern'' dNS form which
admits a non-Abelian generalization. Start with the $(2+1)$d DBI
action:
\be
L= - \sqrt{-\det\left(g_{\mu\nu}+ \frac{1}{g_{YM}}F_{\mu\nu}\right)}
\ee
This action is equivalent to the following action involving a new
independent field $B_\mu$:
\be
\label{abeldual}
L = \half \epsilon^{\mu\nu\lambda}B_\mu F_{\nu\lambda} -
\sqrt{-\det(g_{\mu\nu}+ g_{YM}^2 B_\mu B_\nu)}
\ee
To prove equivalence, simply integrate out $B_\mu$ from the latter
action and recover the former action. 

Now noting that in static gauge, $g_{\mu\nu}=\eta_{\mu\nu} + \del_\mu
X^i \del_\nu X^i$, and making the replacement:
\be
B_\mu \to -\frac{1}{g_{YM}}{\hat D}_\mu X^8 = 
-\frac{1}{g_{YM}}\left(\del_\mu X^8 - B_\mu X_+^8\right)
\ee
we find that the action \eref{abeldual} turns into:
\be
L = \half \epsilon^{\mu\nu\lambda}B_\mu F_{\nu\lambda} -
\sqrt{-\det(\eta_{\mu\nu}+ {\hat D}_\mu X^I {\hat D}_\nu X^I)}
\ee
Hence $SO(8)$ invariance is achieved. It is easily seen that this
subsumes the special (quartic, Abelian) case that we discussed at the
beginning of this section.

The considerations in this section support our conjecture that the
entire non-Abelian D2-brane action can be recast in $SO(8)$ invariant
form, and constitute an important (though long-known) consistency
check on it, since if it works for the non-Abelian case then it must
necessarily work for the Abelian reduction. But to prove the
(non-Abelian) conjecture in general is more difficult, essentially
because the full non-Abelian D-brane action is not yet
known. Having treated the bosonic terms to lowest nontrivial order in
$\alpha'$, we next turn to treatment of the fermionic terms.

\section{Fermionic terms}

The fermionic terms of the action can also be obtained from 10
dimensional supersymmetric gauge theory reduced to three
dimensions. To do this we first need the supersymmetrized DBI action
at ${\alpha'}^2$ level. Then we may reduce the fermionic terms to
three dimension in the same way as we have done for the bosonic
part in a previous section. The aim would be to rewrite the
resulting fermionic terms in $SO(8)$ invariant form.

Let us start with the Abelian case, which has essentially been treated
in the older literature. We will provide a re-derivation which
stresses more explicitly the promotion to $SO(8)$ invariance. This
will be a guide in studying the non-Abelian case. Start with the
following DBI Lagrangian in 10 dimensions \cite{Aganagic:1997zk}:
\be
L= -\sqrt{-\det(\eta_{MN} + F_{MN} - 2\blambda\Gamma_M
\del_N\lambda + \blambda\Gamma^P\del_M\lambda\,\blambda \Gamma_P
\del_N \blambda)}
\ee
Upon dimensional reduction to 3 dimensions, this
reduces to:
\be
-\Bigg(-\left|\begin{matrix}
~\eta_{\mu\nu} + F_{\mu\nu} - 2\blambda \Gamma_\mu\del_\nu\lambda
+\blambda \Gamma^\rho\del_\mu\lambda\,\blambda
\Gamma_\rho\del_\nu\lambda + 
\blambda \Gamma^i\del_\mu\lambda\,\blambda
\Gamma^i\del_\nu\lambda \qquad\qquad & -\del_\mu X^i~\\
\del_\nu X^i - 2\blambda \Gamma_i \del_\nu\lambda &
\eta_{ij}
\end{matrix}\right|\Bigg)^\half
\ee
which can be rewritten as:
\bea
&&-\bigg{[}-\det\Big(\eta_{\mu\nu} + \del_\mu X^i\del_\nu X^i
 - 2\del_\mu X^i \blambda \Gamma^i\del_\nu\lambda
+\blambda \Gamma^i\del_\mu\lambda\blambda \Gamma^i\del_\nu\lambda
+ F_{\mu\nu}
\cr &&\;\;\;\;\;\;\;\;\;\;\;\;\;\;\;\;\;- 2\blambda\Gamma_\mu
\del_\nu\lambda + \blambda\Gamma^\rho\del_\mu\lambda\,\blambda \Gamma_\rho
\del_\nu \blambda\Big)\bigg{]}^{\frac{1}{2}}
\eea

This can now be re-expressed as:
\be
-\sqrt{-\det\left(\tG_{\mu\nu} + D_\mu X^i D_\nu X^i 
+ {\cal F}_{\mu\nu}\right)}
\ee
where:
\bea
{\tilde G}_{\mu\nu} &=& \eta_{\mu\nu}
-2 \blambda \Gamma_{(\mu}\del_{\nu)}\lambda
+\blambda\Gamma^\rho\del_\mu\lambda\,\blambda
\Gamma_\rho\del_\nu\lambda\nn\\
{\cal F}_{\mu\nu} &=& F_{\mu\nu} - 
2 \blambda \Gamma_{[\mu}\del_{\nu]}\lambda
-2 \del_{[\mu}X^i \blambda \Gamma^i \del_{\nu]}\lambda\nn\\
{\hat D}_\mu X^i &=& \del_\mu X^i 
- \blambda \Gamma^i \del_\mu\lambda
\eea
Now following the result in Ref.\cite{Aganagic:1997zk}, the above
action is dual to:
\be
\half \epsilon^{\mu\nu\rho}
\Big(B_\mu-\frac{1}{g_{YM}}\del_\mu X^8\Big) 
{\cal F}_{\nu\rho} -\sqrt{-\det({\tilde G}_{\mu\nu} + {\hat D}_\mu X^I
{\hat D}_\nu X^I)}
\ee
where
\be
{\hat D}_\mu X^8 \equiv \del_\mu X^8 - g_{YM}B_\mu
\ee
and ${\hat D}_\mu X^i=D_\mu X^i,~ i=1,\cdots,7$ which was defined
above.

This expression does not look $SO(8)$ invariant, both for the
Chern-Simons term and the covariant derivative, but we can argue that
in fact both these are $SO(8)$ invariant. First consider the covariant
derivatives. For the $g_{YM} B_\mu$ term we proceed as was explained
for the bosonic case. However, the fermionic term which appears in
${\hat D}_\mu X^i$ is absent in ${\hat D}_\mu X^8$. This seems to pose
a problem for $SO(8)$ invariance. In fact, the quantity:
\be
\Pi_\mu^i = \del X_\mu^i - \blambda\Gamma^i\del_\mu\lambda
\ee
is a supercovariant quantity which occurs in many formulae. So the
question is to understand why
\be
\Pi_\mu^8 = \del_\mu X^8 -\blambda\Gamma^8\del_\mu\lambda
\ee
does not appear. This would be required to form the $SO(8)$ vector
$\Pi_\mu^I$

As explained in Ref.\cite{Aganagic:1996nn}, because we are in static
gauge both with respect to coordinate transformations and
supersymmetries, the fermion $\lambda$ is really a 16-component
fermion descending from the 32-component fermion $\theta$ in the
covariant D-brane formalism. Starting with the original fermionic
variable $\theta$ we define:
\be
\theta_1 = \half (1+\Gamma^8)\theta,\qquad
\theta_2 = \half (1-\Gamma^8)\theta
\ee
(what we call $\Gamma^8$ is referred to as $\Gamma^{11}$ in
Ref.\cite{Aganagic:1996nn}). Then static gauge is chosen by putting
$\theta_2=0$, and rename $\theta_1$ as $\lambda$. Hence:
\be
\Gamma^8 \lambda = \lambda
\ee
It follows that:
\be
\blambda \Gamma^8\del_\mu\lambda = \blambda \del_\mu\lambda = \half
\del_\mu (\blambda\lambda)
\ee
(using the identity $\blambda\chi = {\bar \chi}\lambda$ for
Majorana-Weyl spinors in 10d). Therefore:
\be
\Pi_\mu^8 = \del_\mu \big(X^8 - \half(\blambda\lambda)\big)
\ee
and the second term can be removed by a shift of $X^8$.  This explains
why the covariant derivatives are in fact $SO(8)$ covariant.

For the Chern-Simons term something similar happens. The extra
term compared to the bosonic case is proportional to:
\be
\label{fermterms}
\epsilon^{\mu\nu\rho}
\del_\mu X^8\Big(\blambda \Gamma_\nu\del_\rho\lambda
+ \del_\nu X^i\blambda \Gamma^i \del_\rho\lambda\Big)
\ee
Consider the first term in the above expression. To make it
covariant we would like to write it as:
\be
\epsilon^{\mu\nu\rho}
\del_\mu X^8\,\blambda \Gamma_\nu\del_\rho\lambda=
\epsilon^{\mu\nu\rho}
\del_\mu X^8\,\blambda \Gamma_\nu\Gamma^8\del_\rho\lambda
\to 
\epsilon^{\mu\nu\rho}
\del_\mu X^I\,\blambda \Gamma_\nu\Gamma^I\del_\rho\lambda
\ee
where the first step is an identity (because
$\Gamma^8\lambda=\lambda$) and in the second step we have added a
piece: 
\be 
\epsilon^{\mu\nu\rho}\del_\mu X^i\,\blambda
\Gamma_\nu\Gamma^i\del_\rho\lambda 
\ee 
As we now show, this extra piece is equal to zero, which justifies
adding it to make the above term $SO(8)$ covariant. We have:
\bea 
\epsilon^{\mu\nu\rho}\del_\mu X^i\,\blambda
\Gamma_\nu\Gamma^i\del_\rho\lambda &=& \half
\epsilon^{\mu\nu\rho}\del_\mu X^i\,\blambda
(\Gamma_\nu\Gamma^i- \Gamma^i\Gamma_\nu)
\del_\rho\lambda\nn\\
&=& \frac14 \epsilon^{\mu\nu\rho}\del_\mu X^i\,\del_\rho\Big(\blambda
(\Gamma_\nu\Gamma^i- \Gamma^i\Gamma_\nu)
\lambda\Big)
\eea 
which is zero on integration by parts. (Here we have used the identity
$\blambda \Gamma^{MN}\chi={\bar \chi} \Gamma^{MN}\lambda$).)

Things work similarly for the second term in \eref{fermterms}:
\be
\del_\mu X^8\del_\nu X^i\,\blambda \Gamma^i \del_\rho\lambda
=\del_\mu X^8\del_\nu X^i\,\blambda \Gamma^i\Gamma^8 \del_\rho\lambda
\ee
To make this covariant we need to add:
\be
\half\del_\mu X^i\del_\nu X^j\,\blambda \Gamma^{ij} \del_\rho\lambda
= \frac14\del_\mu X^i\del_\nu X^j\,\del_\rho
\Big(\blambda \Gamma^{ij} \lambda\Big)
\ee
but this is again zero on partial integration. Thus we have shown that
the Abelian fermionic Chern-Simons terms can be written in $SO(8)$
invariant form as:
\be
\epsilon^{\mu\nu\rho}
\del_\mu X^I\,\Big(\blambda \Gamma_\nu\Gamma^I\del_\rho\lambda
+ \half\del_\nu X^J\,\blambda \Gamma^{IJ} \del_\rho\lambda\Big)
\ee

Turning now to the non-Abelian case of interest to us, the relevant
fermionic terms at ${\alpha'}^2$ level in ten dimensional
supersymmetric gauge theory are given by
\cite{{Cederwall:2001td},{Bergshoeff:2001dc}}\footnote{Here we have
not considered terms like
$F\bar{\lambda}\Gamma\lambda\bar{\lambda}\Gamma\lambda$ which from
the string theory point of view are of order of ${\alpha'}^2 g^3$ while the
terms we are considering are of order of ${\alpha'}^2 g^2$.  For details
see \cite{{Cederwall:2001td},{Bergshoeff:2001dc}}.}

\bea
\label{nafermterms}
L_{\rm fer}&=&{\rm Str}\bigg{(}\frac{i}{2}\bar{\lambda}\Gamma^M 
D_M\lambda+\frac{i}{4}
\bar{\lambda}\Gamma_M D^N\lambda F^{MR}F_{RN}-
\frac{i}{8}\bar{\lambda}\Gamma_{MNR} D_S\lambda F^{MN}F^{RS}\cr
&&\;\;\;\;\;\;\;\;-\frac{1}{16}\bar{\lambda}\Gamma^M D^N\lambda\;
\bar{\lambda}\Gamma_N D_M\lambda
\bigg{)}.
\eea

We proceed as follows. First reduce the action to 3
dimensions and then try to rewrite the obtained action in an $SO(8)$
invariant form. Of course one also needs to take the symmetrized trace
{\it Str}. We note however that the first term is easy to deal
with. In fact, dimensionally reducing to three dimensions one gets
\be
\frac{i}{2}
{\rm Tr}\bigg{(}\bar{\lambda}\Gamma^\mu D_\mu\lambda+g_{YM}\bar{\lambda}
\Gamma^i [X^i,\lambda]\bigg{)},
\ee
which can be written as follows:
\be
\label{fermionfinalone}
\frac{i}{2}{\rm Tr}\bigg{(}
\bar{\lambda}\Gamma^\mu D_\mu\lambda+\frac{1}{2}
\bar{\lambda}\Gamma^{IJ} [X^I,X^J,\lambda]
\bigg{)},
\ee
where 
\be
 [X^I,X^J,\lambda]=g_{YM}^I[X^J,\lambda]-g_{YM}^J[X^I,\lambda].
\ee 
The last term in \eref{nafermterms} can also be reduced to three
dimensions, leading to
\bea
&&-\frac{1}{16}{\rm Str}\bigg{(}\bar{\lambda}\Gamma^\mu 
D^\nu\lambda\;\bar{\lambda}\Gamma_\mu D_\nu\lambda
+g_{YM}\bar{\lambda}\Gamma^i D^\nu\lambda\;\bar{\lambda}
\Gamma_\nu [X^i,\lambda]
+g_{YM}\bar{\lambda}\Gamma^\mu 
[X^i,\lambda]\;\bar{\lambda}\Gamma^i D_\mu\lambda\cr
&&\;\;\;\;\;\;\;\;\;+g^2_{YM}\bar{\lambda}\Gamma^i [X^j,
\lambda]\;\bar{\lambda}\Gamma^j [X^i,\lambda]\bigg{)}
\eea
Using our notation the above action can be recast in the following
$SO(8)$ invariant form
\bea
\label{fermionfinaltwo}
&&-\frac{1}{16}{\rm Str}\bigg{(}\bar{\lambda}\Gamma^\mu
D^\nu\lambda\;\bar{\lambda}\Gamma_\mu D_\nu\lambda
+\frac{1}{4}g^2_{YM}\bar{\lambda}\Gamma^{IJ}
[X^K,X^L,\lambda]\;\bar{\lambda}\Gamma^{KL} [X^I,X^J,\lambda]
\cr
&&\;\;\;\;\;\;\;\;\;\;\;\;\;\; 
+\frac{1}{2}\bar{\lambda}\Gamma^\mu
[X^I,X^J,\lambda]\;\bar{\lambda}\Gamma^{IJ} D_\mu\lambda
+\frac{1}{2}\bar{\lambda}\Gamma^{IJ}
D^\nu\lambda\;\bar{\lambda}\Gamma_\nu [X^I,X^J,\lambda]
\bigg{)}
\eea
Of course we still need to take the symmetrized trace {\rm Str}.

The second and third terms in \eref{nafermterms} are more involved. For
these terms it is useful to first expand the {\rm Str} (of course at
the end we will again rewrite the action in terms of {\rm Str}). Doing
so, we get
\bea
&&{\rm Str}\bigg{(}\frac{i}{4}
\bar{\lambda}\Gamma_M D^N\lambda F^{MR}F_{RN}-
\frac{i}{8}\bar{\lambda}\Gamma_{MNR} D_S\lambda F^{MN}F^{RS}\bigg{)}\cr
&&=\frac{1}{3!}{\rm Tr}\bigg{[}\bigg{(}\frac{i}{4}
\bar{\lambda}\Gamma_M D^N\lambda F^{MR}F_{RN}-
\frac{i}{8}\bar{\lambda}\Gamma_{MNR} D_S\lambda F^{MN}F^{RS}\bigg{)}\cr
&&\;\;\;\;\;\;\;\;\;\;\;\;\;\bigg{(}\frac{i}{4}
\bar{\lambda}\Gamma_M D^N\lambda 
F_{RN} F^{MR}-\frac{i}{8}\bar{\lambda}\Gamma_{MNR} 
D_S\lambda F^{RS}F^{MN}\bigg{)}\cr
&&\;\;\;\;\;\;\;\;\;\;\;\;\;+{\rm 4\; 
more\; pairs\; obtained\; from \; permutations\; 
of}\; F_{MN}\;{\rm and}\; \lambda
\bigg{]}.
\eea
We note, however, that to reduce and convert the obtained action to
the $SO(8)$ invariant terms we do not need the four extra pairs coming
from the permutations. As soon as we get the $SO(8)$ invariant from of
the first two pairs, the others can be obtained by an obvious
permutation of $\lambda$'s and $\hat{D}X^J$'s. So in what follows we
just concentrate on the first two pairs.

Reducing the above part of the fermionic action from the first two
pairs, one finds:
\bea
&&\frac{1}{3!}{\rm Tr}\bigg{[}\bigg{(}\frac{i}{4}
\bigg{\{}\frac{1}{g_{YM}^2}\bar{\lambda}\Gamma_\mu D^\nu\lambda\; 
F^{\mu\rho}F_{\rho\nu}-
\bar{\lambda}\Gamma_\mu D^\nu\lambda\; 
D^\mu X^l D_\nu X^l-\frac{1}{g_{YM}}
\bar{\lambda}\Gamma^i D^\nu\lambda\; D^\rho X^i F_{\rho\nu}\cr
&&\;\;\;\;\;\;-~g_{YM}\bar{\lambda}\Gamma^i D^\nu\lambda\; X^{il} D_\nu X^l
+\bar{\lambda}\Gamma_\mu [X^j,\lambda]\; F^{\mu\rho} D_\rho X^j\cr
&&\;\;\;\;\;\;-~g_{YM}\bar{\lambda}\Gamma^i [X^j,\lambda]\; 
D^\rho X^i D_\rho X^j
+g_{YM}^2\bar{\lambda}\Gamma_\mu [X^j,\lambda] D^\mu X^l X^{lj}\cr
&&\;\;\;\;\;\;+~g_{YM}^3\bar{\lambda}\Gamma^i [X^j,\lambda] 
X^{il} X^{lj}\bigg{\}}\cr
&&-\frac{i}{8}\bigg{\{}\frac{1}{g_{YM}^2}\bar{\lambda}
\Gamma_{\mu\nu\rho} D_\sigma\lambda\; F^{\mu\nu}F^{\rho\sigma}
+\bar{\lambda}\Gamma_{\mu\nu\rho} [X^k,\lambda]\; F^{\mu\nu}D^\rho X^k
-\frac{1}{g_{YM}}\bar{\lambda}\Gamma_{\mu\nu l} D_\sigma\lambda\; 
F^{\mu\nu}D^\sigma X^l\cr
&&\;\;\;\;\;\;\;\;+~g_{YM}\bar{\lambda}\Gamma_{\mu\nu l} 
[X^k,\lambda]\; F^{\mu\nu}X^{lk}
+\frac{2}{g_{YM}}\bar{\lambda}\Gamma_{\mu j\rho} 
D_\sigma\lambda\; D^{\mu} X^j F^{\rho\sigma}\cr
&&\;\;\;\;\;\;\;\;+~2g_{YM}\bar{\lambda}\Gamma_{\mu j\rho} 
[X^k,\lambda]\; D^{\mu}X^jD^{\rho}X^k
-2\bar{\lambda}\Gamma_{\mu jl} D_\sigma\lambda\; 
D^{\mu}X^jD^{\sigma}X^l\cr
&&\;\;\;\;\;\;\;\;+~2g^2_{YM}\bar{\lambda}\Gamma_{\mu j l} 
[X^k,\lambda]\; D^{\mu}X^j X^{lk}
+\bar{\lambda}\Gamma_{ij\rho} D_\sigma\lambda\; X^{ij}F^{\rho\sigma}\cr
&&\;\;\;\;\;\;\;\;+~g^2_{YM}\bar{\lambda}\Gamma_{ij\rho} 
[X^k,\lambda]\; X^{ij}D^{\rho}X^k
-g_{YM}\bar{\lambda}\Gamma_{ijl} D_\sigma\lambda\; 
X^{ij}D^{\sigma}X^l\cr
&&\;\;\;\;\;\;\;\;+~g^3_{YM}\bar{\lambda}\Gamma_{ijl} 
[X^k,\lambda]\; X^{ij}X^{lk}
\bigg{\}}\bigg{)}+\bigg{(}{\rm the\; same\; terms\; with}\; 
F \leftrightarrow F\bigg{)}+\cdots\bigg{]}.
\eea

Now the task is to rewrite these terms in $SO(8)$ invariant form. To
do this, following the procedure of the previous section we should
first dualize $F$ to $B$ field and then use the shift symmetry to
replace $B$ by $\hat{D}X^8$. Using the properties of 3D gamma matrices
and dropping terms which are zero on shell\footnote{More precisely we
have $\epsilon^{\mu\nu\rho}\gamma_\rho=\gamma^{\mu\nu}$. Moreover one
will drop all terms involving ${\alpha'}^2
(\gamma_\mu\partial^\mu\lambda+g_{YM}\gamma^i[X^i,\lambda])$.}  one
arrives at
\bea
\label{fermionfinalthree}
&&\frac{i}{8}{\rm Str}\bigg{[}
2\bar{\lambda}\Gamma_\mu \Gamma^{IJ}D_\nu\lambda \hat{D}^\mu X^I\hat{D}^\nu X^J
-2\bar{\lambda}\Gamma_\mu D^\nu\lambda \hat{D}^\mu X^I\hat{D}_\nu X^I
\\
&&\;\;\;\;\;\;\;\;\;
+\bar{\lambda} \Gamma^{IJKL}D_\nu\lambda\; X^{IJK}\hat{D}^\nu X^L
-\bar{\lambda} \Gamma^{IJ}D_\nu\lambda\; X^{IJK}\hat{D}^\nu X^K
\cr&&\;\;\;\;\;\;\;\;\;
-2\bar{\lambda} \Gamma^{IJ}[X^J,X^K,\lambda] \hat{D}^\mu X^I\hat{D}_\mu X^K
-2\bar{\lambda}[X^I,X^J,\lambda] \hat{D}^\mu X^I\hat{D}_\mu X^J
\cr&&\;\;\;\;\;\;\;\;\;
-2\bar{\lambda}\Gamma^{\mu\nu}[X^I,X^J,\lambda] \hat{D}_\mu X^I\hat{D}_\nu X^J
-2\bar{\lambda}\Gamma_{\mu\nu}\Gamma^{IJ}[X^J,X^K,\lambda] \hat{D}^\mu X^I\hat{D}^\nu X^K
\cr&&\;\;\;\;\;\;\;\;\;
+\bar{\lambda}\Gamma_\mu\Gamma^{IJ}[X^K,X^L,\lambda] \hat{D}^\mu X^IX^{JKL}
-\bar{\lambda}\Gamma_\mu[X^I,X^J,\lambda] \hat{D}^\mu X^KX^{IJK}
\cr&&\;\;\;\;\;\;\;\;\;
-\frac{1}{3}\bar{\lambda}\Gamma_\mu\Gamma^{IJKL}[X^L,X^M,\lambda] X^{IJK} \hat{D}^\mu X^M 
-\bar{\lambda}\Gamma_\mu\Gamma^{IJ}[X^K,X^L,\lambda] X^{IJK}\hat{D}^\mu X^L
\cr&&\;\;\;\;\;\;\;\;\;
-\frac{1}{6}\bar{\lambda}\Gamma^{IJKL}[X^M,X^N,\lambda] X^{IJL}X^{KMN}
-\frac{1}{2}\bar{\lambda}\Gamma^{IJ}[X^K,X^L,\lambda] X^{IJM}X^{KLM}
\bigg{]}.\nonumber
\eea

To summarize this section, we have found the $SO(8)$ invariant
fermionic terms to lowest nontrivial order in $\alpha'$ and they are
contained in the sum of
Eqs.(\ref{fermionfinalone}),(\ref{fermionfinaltwo}),
(\ref{fermionfinalthree}).

\section{A conjecture}

A striking aspect of our result for higher derivative corrections is
that it can be written in a form that only uses basic objects of
3-algebras: the covariant derivative on scalars and fermions, and the
triple product $[X^I,X^J,X^K]$ and $[X^I,X^J,\lambda]$. To leading order in derivatives we have
written the complete answer, for both bosons and fermions, and we
expect it is maximally supersymmetric (though we did not prove that
here).

Given this situation, it seems reasonable to speculate that the same
derivative corrections are relevant to all 3-algebras with maximal
supersymmetry, regardless of their signature. For Euclidean signature,
this in fact only includes just one theory besides the ones we were
considering, namely the Bagger-Lambert $A_4$
theory\cite{Bagger:2007vi}\footnote{For arbitrary signature
it is possible to construct more such algebras. In particular,
algebras with $(2,p)$ signature have been classified in
\cite{deMedeiros:2008bf}. We would like to thank Jose Figueroa-O'
Farrill for a comment on this point.}. Thus we conjecture that the
action in
Eqs.(\ref{bosonfinal}),(\ref{fermionfinalone}),(\ref{fermionfinaltwo}),
(\ref{fermionfinalthree}) also embodies the derivative corrections to
the Euclidean 3-algebra $A_4$ theory.

It may legitimately be argued that there is no concrete test of this
conjecture given that we do not presently know how to compute
derivative corrections to the membrane field theory starting from
M-theory. However an important test in our opinion will be whether the
higher-derivative theory we have constructed is really maximally
supersymmetric. Since our Lagrangian inherits its entire structure
from ${\cal N}=8$ SYM, this must surely be the case. Assuming
supersymmetry can be proved, it is most likely that the proof will
rely only on abstract 3-algebra properties and therefore will go
through in the same way for the $A_4$ theory.

\section{Conclusions}

In this paper we have shown that the world-volume theory of multiple
D2-branes in string theory, including both the ${\cal N}=8$ SYM part
as well as the leading (bosonic and fermionic) higher derivative
corrections, is equivalent by a dNS duality to a derivative-corrected
Lorentzian 3-algebra theory. This generalizes the result in
\cite{Ezhuthachan:2008ch} to incorporate $\alpha'$ corrections. We see
no obstacle in principle to extending this to any finite order in
$\alpha'$ as long as the D2-brane action is known to that order. 

The result has the elegant feature that it depends only on 3-algebra
quantities: the 3-bracket and covariant derivative. We have
conjectured that it has more general significance than the context in
which we have derived it. Extended supersymmetric CFT's in 3
dimensions appear to all depend on the 3-algebra structure (although
if ${\cal N}<8$ then some of the original 3-algebra assumptions need
to be relaxed\cite{Schnabl:2008wj,Bagger:2008se}). Our results can be
extended in a straightforward manner only to the Euclidean $A_4$
3-algebra but in the future, with extra work, it should be possible to
extend them at least to the ${\cal N}=6$ case.\\[4mm]

{\bf Note added:} While this article was being prepared
Ref.\cite{Iengo:2008cq} appeared on the arXiv, in which a non-linear
theory for multiple M2-branes has been proposed. Earlier papers that
might be related are \cite{{Li:2008ya},{Kluson:2008nw},{Bonelli:2008kh}}.

\acknowledgments{We are grateful to Bobby Ezhuthachan, Mohammad Garousi, 
Neil Lambert, Costis Papageorgakis and Ashoke Sen for helpful
discussions.  M.A. would like to thank TIFR and the organizers of the
Monsoon Workshop on String Theory at TIFR, where this work was
initiated, for very warm hospitality. M. A. is also supported in part
by the Iranian TWAS chapter at ISMO. S.M. is grateful to CERN, Geneva
for hospitality during the completion of this work. The work of
S.M. was supported in part by a J.C. Bose Fellowship of the Government
of India.}


\begin{thebibliography}{10}

\bibitem{Bagger:2006sk}
  J.~Bagger and N.~Lambert, ``Modeling multiple M2's,'' Phys.\ Rev.\ D
  {\bf 75}, 045020 (2007) [arXiv:hep-th/0611108].  
  PHRVA,D75,045020;

\bibitem{Bagger:2007jr}
  J.~Bagger and N.~Lambert,
  ``Gauge Symmetry and Supersymmetry of Multiple M2-Branes,''
  Phys.\ Rev.\  D {\bf 77}, 065008 (2008)
  [arXiv:0711.0955 [hep-th]].

\bibitem{Bagger:2007vi}
  J.~Bagger and N.~Lambert,
  ``Comments On Multiple M2-branes,''
  JHEP {\bf 0802}, 105 (2008)
  [arXiv:0712.3738 [hep-th]].

\bibitem{Gustavsson:2007vu}
  A.~Gustavsson,
  ``Algebraic structures on parallel M2-branes,''
  arXiv:0709.1260 [hep-th].

\bibitem{Gran:2008vi}
  U.~Gran, B.~E.~W.~Nilsson and C.~Petersson,
  ``On relating multiple M2 and D2-branes,''
  arXiv:0804.1784 [hep-th].


\bibitem{Gomis:2008uv}
  J.~Gomis, G.~Milanesi and J.~G.~Russo,
  ``Bagger-Lambert Theory for General Lie Algebras,''
  JHEP {\bf 0806}, 075 (2008)
  [arXiv:0805.1012 [hep-th]].

\bibitem{Benvenuti:2008bt}
  S.~Benvenuti, D.~Rodriguez-Gomez, E.~Tonni and H.~Verlinde,
  ``N=8 superconformal gauge theories and M2 branes,''
  arXiv:0805.1087 [hep-th].

\bibitem{Ho:2008ei}
  P.~M.~Ho, Y.~Imamura and Y.~Matsuo,
  ``M2 to D2 revisited,''
  JHEP {\bf 0807}, 003 (2008)
  [arXiv:0805.1202 [hep-th]].

\bibitem{Bandres:2008kj}
  M.~A.~Bandres, A.~E.~Lipstein and J.~H.~Schwarz,
  ``Ghost-Free Superconformal Action for Multiple M2-Branes,''
  arXiv:0806.0054 [hep-th].

\bibitem{Gomis:2008be}
  J.~Gomis, D.~Rodriguez-Gomez, M.~Van Raamsdonk and H.~Verlinde,
  ``Supersymmetric Yang-Mills Theory From Lorentzian Three-Algebras,''
  arXiv:0806.0738 [hep-th].

\bibitem{Mukhi:2008ux}
  S.~Mukhi and C.~Papageorgakis,
  ``M2 to D2,''
  JHEP {\bf 0805}, 085 (2008)
  [arXiv:0803.3218 [hep-th]].

\bibitem{Honma:2008un}
Y.~Honma, S.~Iso, Y.~Sumitomo, and S.~Zhang, ``Janus field theories from
  multiple M2 branes,''  arXiv:0805.1895.

\bibitem{Ezhuthachan:2008ch}
  B.~Ezhuthachan, S.~Mukhi and C.~Papageorgakis,
  ``D2 to D2,''
  JHEP {\bf 0807}, 041 (2008)
  [arXiv:0806.1639 [hep-th]].
  
  
\bibitem{Bergshoeff:2008ix}
  E.~A.~Bergshoeff, M.~de Roo, O.~Hohm and D.~Roest,
  ``Multiple Membranes from Gauged Supergravity,''
  arXiv:0806.2584 [hep-th].

\bibitem{Nicolai:2003bp}
  H.~Nicolai and H.~Samtleben,
  ``Chern-Simons vs. Yang-Mills gaugings in three dimensions,''
  Nucl.\ Phys.\  B {\bf 668}, 167 (2003)
  [arXiv:hep-th/0303213].

\bibitem{deWit:2003ja}
  B.~de Wit, I.~Herger and H.~Samtleben,
  ``Gauged locally supersymmetric D = 3 nonlinear sigma models,''
  Nucl.\ Phys.\  B {\bf 671}, 175 (2003)
  [arXiv:hep-th/0307006].

\bibitem{deWit:2004yr}
  B.~de Wit, H.~Nicolai and H.~Samtleben,
  ``Gauged supergravities in three dimensions: A panoramic overview,''
  arXiv:hep-th/0403014.
  
\bibitem{Tseytlin:1997csa}
  A.~A.~Tseytlin,
  ``On non-Abelian generalisation of the Born-Infeld action in string
  theory,''
  Nucl.\ Phys.\  B {\bf 501}, 41 (1997)
  [arXiv:hep-th/9701125].
  
  
\bibitem{Schnabl:2008wj}
  M.~Schnabl and Y.~Tachikawa,
  ``Classification of N=6 superconformal theories of ABJM type,''
  arXiv:0807.1102 [hep-th].

\bibitem{Bagger:2008se}
  J.~Bagger and N.~Lambert,
  ``Three-Algebras and N=6 Chern-Simons Gauge Theories,''
  arXiv:0807.0163 [hep-th].


\bibitem{Bergshoeff:2008bh}
  E.~A.~Bergshoeff, O.~Hohm, D.~Roest, H.~Samtleben and E.~Sezgin,
  ``The Superconformal Gaugings in Three Dimensions,''
  arXiv:0807.2841 [hep-th].


\bibitem{Townsend:1995af}
  P.~K.~Townsend,
  ``D-branes from M-branes,''
  Phys.\ Lett.\  B {\bf 373}, 68 (1996)
  [arXiv:hep-th/9512062].

  
\bibitem{Schmidhuber:1996fy}
  C.~Schmidhuber,
  ``D-brane actions,''
  Nucl.\ Phys.\  B {\bf 467} (1996) 146
  [arXiv:hep-th/9601003].

\bibitem{Aganagic:1997zk}
  M.~Aganagic, J.~Park, C.~Popescu and J.~H.~Schwarz,
  ``Dual D-brane actions,''
  Nucl.\ Phys.\  B {\bf 496}, 215 (1997)
  [arXiv:hep-th/9702133].

\bibitem{Aganagic:1996nn}
  M.~Aganagic, C.~Popescu and J.~H.~Schwarz,
  ``Gauge-invariant and gauge-fixed D-brane actions,''
  Nucl.\ Phys.\  B {\bf 495}, 99 (1997)
  [arXiv:hep-th/9612080].



\bibitem{Cederwall:2001td}
  M.~Cederwall, B.~E.~W.~Nilsson and D.~Tsimpis,
  ``D = 10 super-Yang-Mills at ${\cal O}({\alpha'\,}^2)$,''
  JHEP {\bf 0107}, 042 (2001)
  [arXiv:hep-th/0104236].


\bibitem{Bergshoeff:2001dc}
  E.~A.~Bergshoeff, A.~Bilal, M.~de Roo and A.~Sevrin,
  ``Supersymmetric non-Abelian Born-Infeld revisited,''
  JHEP {\bf 0107}, 029 (2001)
  [arXiv:hep-th/0105274].


\bibitem{deMedeiros:2008bf}
  P.~de Medeiros, J.~M.~Figueroa-O'Farrill and E.~Mendez-Escobar,
  ``Metric Lie 3-algebras in Bagger-Lambert theory,''
  arXiv:0806.3242 [hep-th].


\bibitem{Iengo:2008cq}
  R.~Iengo and J.~G.~Russo,
  ``Non-linear theory for multiple M2 branes,''
  arXiv:0808.2473 [hep-th].

\bibitem{Li:2008ya}
  T.~Li, Y.~Liu and D.~Xie,
  ``Multiple D2-Brane Action from M2-Branes,''
  arXiv:0807.1183 [hep-th].
  
 

\bibitem{Kluson:2008nw}
  J.~Kluson,
  ``D2 to M2 Procedure for D2-Brane DBI Effective Action,''
  arXiv:0807.4054 [hep-th].




\bibitem{Bonelli:2008kh}
  G.~Bonelli, A.~Tanzini and M.~Zabzine,
  ``Topological branes, p-algebras and generalized Nahm equations,''
  arXiv:0807.5113 [hep-th].







\end{thebibliography}
\end{document}